\documentclass{new_tlp}
\usepackage{mathptmx}
\usepackage[ruled,linesnumbered,vlined,noline]{algorithm2e}
\usepackage{xspace}
\usepackage{url}
\usepackage{tikz}

\newcommand{\wasp}{\textsc{wasp}\xspace}
\newcommand{\hclasp}{\textsc{hclasp}\xspace}
\newcommand{\clasp}{\textsc{clasp}\xspace}
\newcommand{\gringo}{\textsc{gringo}\xspace}
\newcommand{\clingo}{\textsc{clingo}\xspace}

\newcommand{\minisat}{\textsc{minisat}\xspace}
\newcommand{\coherent}{\textsc{coherent}\xspace}
\newcommand{\inco}{\textsc{incoherent}\xspace}

\newcommand{\heuristic}{\texttt{\detokenize{_heuristic}}\xspace}

\def\naf{\ensuremath{\raise.17ex\hbox{\ensuremath{\scriptstyle\mathtt{\sim}}}}\xspace}

\def\L{\ensuremath{\mathcal{L}}\xspace}

\newcommand{\GP}[1]{\ensuremath{Ground(#1)}\xspace}
\newcommand{\HB}[1]{\ensuremath{B_{#1}}\xspace}
\newcommand{\HU}[1]{\ensuremath{U_{#1}}\xspace}

\newcommand\bcmdtab{\noindent\bgroup\tabcolsep=0pt%
	\begin{tabular}{@{}p{10pc}@{}p{20pc}@{}}}
	\newcommand\ecmdtab{\end{tabular}\egroup}

\title[The External Interface for Extending WASP]
{Technical Note\\ The External Interface for Extending WASP \thanks{The paper has been partially supported by the Italian Ministry for Economic Development (MISE) under project ``PIUCultura -- Paradigmi Innovativi per l'Utilizzo della Cultura'' (n. F/020016/01-02/X27), and under project ``Smarter Solutions in the Big Data World (S2BDW)'' (n. F/050389/01-03/X32) funded within the call ``HORIZON2020'' PON I\&C 2014-2020. Authors are grateful to Mario Alviano, Bernardo Cuteri, Philip Gasteiger, Nicola Leone, Benjamin Musitsch, Peter Sch\"uller, and Kostyantyn Shchekotykhin for their suggestions.}}

\author[Carmine Dodaro and Francesco Ricca]
{CARMINE DODARO\\
	DIBRIS, University of Genova, Genova, Italy\\
	\email{dodaro@dibris.unige.it}\\
\and FRANCESCO RICCA\\
DEMACS, University of Calabria, Rende, Italy\\
\email{ricca@mat.unical.it}
}

\jdate{March 2003}
\pubyear{2003}
\pagerange{\pageref{firstpage}--\pageref{lastpage}}
\doi{S1471068401001193}

\begin{document}
	
	\label{firstpage}
	
	\maketitle
	
	\begin{abstract}
	Answer set programming (ASP) is a successful declarative formalism for knowledge representation and reasoning. 
	The evaluation of ASP programs is nowadays based on the Conflict-Driven Clause Learning (CDCL) backtracking search algorithm.
	%
	Recent work suggested that the performance of CDCL-based implementations can be considerably improved on specific benchmarks by extending their solving capabilities with custom heuristics and propagators. 
	However, embedding such algorithms into existing systems requires expert knowledge of the internals of ASP implementations. The development of effective solver extensions can be made easier by providing suitable programming interfaces. 
	In this paper, we present the interface for extending the CDCL-based ASP solver \wasp.
	The interface is both \textit{general}, i.e. it can be used for providing either new branching heuristics and propagators, and \textit{external}, i.e. the implementation of new algorithms requires no internal modifications of \wasp.
	Moreover, we review the applications of the interface witnessing it can be successfully used to extend \wasp for solving effectively hard instances of both real-world and synthetic problems.
	Under consideration in Theory and Practice of Logic Programming (TPLP).
	\end{abstract}
	
		
	\section{Introduction}
	Answer set programming (ASP) is a declarative formalism for knowledge representation and reasoning based on stable model semantics \cite{DBLP:journals/ngc/GelfondL91,DBLP:journals/cacm/BrewkaET11}.
	ASP has been applied for solving complex problems in several areas, including Artificial Intelligence~\cite{DBLP:conf/lpnmr/BalducciniGWN01,DBLP:journals/aicom/GarroPR06,DBLP:conf/rr/DodaroLNR15}, Bioinformatics~\cite{DBLP:journals/tplp/ErdemO15,DBLP:journals/tplp/KoponenOJS15},  Hydroinformatics~\cite{DBLP:journals/logcom/GavanelliNP15}, Databases~\cite{DBLP:journals/dke/MarileoB10,DBLP:journals/tplp/MannaRT15,DBLP:journals/tplp/MannaRT13}, and Scheduling~\cite{DBLP:conf/aiia/AlvianoDM17,DBLP:journals/fuin/AbseherGMSW16,DBLP:conf/lpnmr/DodaroM17}, to mention a few; see \cite{DBLP:journals/aim/ErdemGL16} for a detailed survey on ASP applications.
	The success of ASP is due to the combination of its high knowledge-modeling power with robust solving technology~\cite{DBLP:journals/ai/GebserKS12,DBLP:conf/jelia/MarateaPR12,DBLP:conf/lpnmr/AlvianoDLR15,DBLP:conf/lpnmr/GebserMR15,DBLP:conf/aaai/GebserMR16,DBLP:journals/aim/LierlerMR16,DBLP:journals/jair/GebserMR17,DBLP:conf/lpnmr/AlvianoCDFLPRVZ17}.

	State-of-the-art ASP systems are usually based on the ``ground+solve'' approach~\cite{DBLP:journals/aim/KaufmannLPS16}, in which a \textit{grounder} transforms the input program (containing variables) in an equivalent variable-free one, whose stable models are subsequently computed by the \textit{solver}.
	The computation of stable models is usually performed applying a variant of the Conflict-Driven Clause Learning (CDCL) backtracking search algorithm~\cite{DBLP:journals/tc/Marques-SilvaS99,DBLP:conf/iccad/ZhangMMM01}.
	Effective CDCL implementations require the combination of several features, including \textit{choice heuristic} and \textit{propagators}.
	Recently, it has been suggested that the performance of CDCL-based solvers can be considerably improved on specific benchmarks by adding domain-specific choice heuristics~\cite{InvitedFriedrich15} and propagators~\cite{DBLP:conf/aaai/JanhunenTT16}.
%
	%
	However, extending in an effective way existing solvers with new algorithms is not obvious, because it requires in depth knowledge of the internals of the implementations, which are nowadays very optimized and sophisticated.

	In this paper, we provide a practical contribution in the aforementioned context, in particular we present the external programming interface of the ASP solver \wasp~\cite{DBLP:conf/cilc/DodaroAFLRS11,DBLP:conf/lpnmr/AlvianoDFLR13,DBLP:conf/lpnmr/AlvianoDLR15}, whose idea is to simplify the integration of custom heuristics and propagators in the solver.	
	In particular, it offers multi-language support including \textit{python} and \textit{perl} languages that require no modifications to the solver, as well as \textit{C++} for performance-oriented implementations.
	The interface was partially described in \cite{DBLP:journals/tplp/DodaroGLMRS16} and used in the literature to embed domain-specific heuristics for two industrial problems proposed by Siemens, namely Partner Units and Combined Configuration.
	More recently, the interface has been also used to implement custom propagators as solution to the \textit{grounding bottleneck} problem in three benchmarks, namely Stable Marriage, Packing and Natural Language Understanding~\cite{DBLP:journals/tplp/CuteriDRS17}.
	In particular, propagators were used to replace a small set of constraints causing a grounding blow-up of the program, and thus making the usage of plain ``ground+solve'' approach not viable. 

	This paper is organized as follows: 
		In Section~\ref{sec:preliminaries}, we recall syntax and semantics of propositional ASP programs and contemporary solving techniques.
		In Section~\ref{sec:interface} we present the interface for extending the solving capabilities of \wasp by adding new propagators and by modifying the choice heuristic.
		Subsequently, we show the usage of the interface by presenting three examples. In the first example, we realize new propagators as a tool for avoiding the grounding bottleneck while solving the Stable Marriage problem (Section~\ref{sec:example:stablemarriage}).
		In the second example, we show how to obtain a naive solver for Constraint Answer Set Programming (CASP)~\cite{DBLP:conf/iclp/BaseliceBG05} (Section~\ref{sec:example:casp}).
		The third one shows how the interface can be used for implementing a well-known general purpose choice heuristic (Section~\ref{sec:example:vsids}).
		Then, we review the successful applications of the interface in the literature, discussing its impact in improving the performance of \wasp (Section~\ref{sec:application}). Finally, after discussing related works in Section~\ref{sec:related}, we draw the conclusion.
	
	\section{Preliminaries}\label{sec:preliminaries}
	This section recalls syntax and semantics of propositional Answer Set Programming (ASP) programs and contemporary solving techniques.
	More detailed descriptions and a more formal account of Answer Set Programming (ASP), including the features of the language employed in this paper, can be found in~\cite{baral2003,DBLP:journals/cacm/BrewkaET11,DBLP:series/synthesis/2012Gebser,DBLP:journals/aim/Lifschitz16,DBLP:journals/aim/JanhunenN16}.
	Hereafter, we assume the reader is familiar with logic programming conventions.
	
	\subsection{ASP Syntax and Semantics}
	
	\paragraph{Syntax.}
	An ASP program $\Pi$ is a finite set of rules of the form:
	\begin{equation}\label{eq:rule}
	a_1 \lor \ldots \lor a_n \leftarrow b_1, \ldots, b_j, \naf b_{j+1}, \ldots, \naf b_m
	\end{equation}
	where $a_1,\ldots,a_n,b_1,\ldots,b_m$ are atoms and $n\geq 0,$
	$m\geq j\geq 0$.
	In particular, an \emph{atom} is an expression of the form $p(t_1, \ldots, t_k)$, where $p$ is a predicate symbol and $t_1, \ldots, t_k$ are \emph{terms}. 
	Terms are alphanumeric strings, and are divided into variables and constants.
	According to Prolog's convention, only variables start with an uppercase letter.
	A \emph{literal} is an atom $a_i$ (positive) or its negation $\naf a_i$ (negative), where $\naf$ denotes the \emph{negation as failure}.
	Given a literal $\ell$, let $\overline{\ell}$ denote the complement of $\ell$, i.e. $\overline{a}=\naf a$ and $\overline{\naf a}=a$, for an atom $a$.
	For a set of literals $S$, let $\overline{S}=\{\overline{\ell} \mid \ell \in S\}$.
	Given a rule $r$ of the form (\ref{eq:rule}), the disjunction $a_1 \lor \ldots \lor a_n$ is the {\em head} of $r$, while $b_1,\ldots,b_j, \naf b_{j+1}, \ldots, \naf b_m$ is the {\em body} of $r$, of which $b_1,\ldots,b_j$ is the {\em positive body}, and $\naf b_{j+1}, \ldots, \naf b_m$ is the {\em negative body} of $r$.
	A rule $r$ of the form (\ref{eq:rule}) is called a \textit{fact} if $m=0$ and a \textit{constraint} if $n=0$.
	Given a rule $r$ of the form (\ref{eq:rule}), $\mathit{H}(r)$ and $\mathit{B}(r)$ denote the set of atoms appearing in the head and in the body of $r$, respectively.
	An object (atom, rule, etc.) is called {\em ground} or {\em propositional}, if it contains no variables. 
	Rules and programs are \textit{positive} if they contain no negative literals, and \textit{general} otherwise.
	Given a program $\Pi$, let the \emph{Herbrand Universe} \HU{\Pi} be the
	set of all constants appearing in $\Pi$ and the \emph{Herbrand Base}
	\HB{\Pi} be the set of all possible ground atoms which can be constructed
	from the predicate symbols appearing in $\Pi$ with the constants of \HU{\Pi}.
	Given a rule $r$, \GP{r} denotes the
	set of rules obtained by applying all possible substitutions $\sigma$
	from the variables in $r$ to elements of \HU{\Pi}. Similarly, given a
	program $\Pi$, the {\em ground instantiation} \GP{\Pi} of $\Pi$
	is the set \( \bigcup_{r \in \Pi} \GP{r} \).
	Given a program $\Pi$, $\mathit{At}(\Pi)$ denotes the set of atoms occurring in $\Pi$.
	
	\paragraph{Semantics.}
	For every program $\Pi$, its stable models are defined using its ground
	instantiation \GP{\Pi} in two steps: First stable models of positive
	programs are defined, then a reduction of general programs to positive
	ones is given, which is used to define stable models of general
	programs.
	
	A set $L$ of ground literals is said to be {\em consistent} if, for every
	literal $\ell \in L$, its negated literal $\overline{\ell}$ is not contained in
	$L$. 
	Given a set of ground literals $L$, $L^+ \subseteq L$ denotes 
	the set of positive literals in $L$.
	An interpretation $I$ for $\Pi$ is a consistent set of ground literals
	over atoms in $\HB{\Pi}$.
	A ground literal $\ell$ is {\em true} w.r.t.\ $I$ if $\ell\in I$;
	$\ell$ is {\em false} w.r.t.\ $I$ if its negated literal is in $I$;
	$\ell$ is {\em undefined} w.r.t.\ $I$ if it is neither true nor false w.r.t.\ $I$.
	A rule $r$ is satisfied w.r.t. $I$ if one of the atoms in the head is true w.r.t.\ $I$ or one of the literals in the body is false w.r.t.\ $I$.
	A constraint $c$ is said to be \textit{violated} by an interpretation $I$ if all literals in the body of $c$ are true.
	An interpretation $I$ is {\em total} if, for each atom $a$ in $\HB{\Pi}$,
	either $a$ or $\naf a$ is in $I$ (i.e., no atom in $\HB{\Pi}$ is undefined w.r.t.\ $I$).
	Otherwise, it is \textit{partial}.
	A total interpretation $M$ is a {\em model} for $\Pi$
	if, for every $r \in \GP{\Pi}$, at
	least one literal in the head of $r$ is true w.r.t.\ $M$ whenever all literals in the
	body of $r$ are true w.r.t.\ $M$.
	A model $M$ is a {\em stable model}
	for a positive program $\Pi$ if $M^+ \subseteq X^+$, for each model $X$ of $\Pi$ .
	
	The {\em reduct} of a general ground program $\Pi$ w.r.t.\ an interpretation $M$ is the positive ground program $\Pi^M$, obtained from $\Pi$ by (i) deleting all rules
	$r \in \Pi$ whose negative body is false w.r.t.\ $M$ and (ii)
	deleting the negative body from the remaining rules.
	A stable model (or answer set) of $\Pi$ is a model $M$ of $\Pi$ such
	that $M$ is a stable model of $\GP{\Pi}^M$. 
	We denote by $\mathit{SM}(\Pi)$ the set of all stable models of $\Pi$,
	and call $\Pi$ \textit{coherent} if $\mathit{SM}(\Pi) \neq \emptyset$, \textit{incoherent} otherwise.
	
	\paragraph{Support.}
	Given a model $M$ for a ground program $\Pi$, 
	we say that a ground atom $a \in M$ is {\em supported}
	with respect to $M$ if there exists a \emph{supporting} rule $r\in \Pi$
	such that $a$ is the only true atom w.r.t. $M$ in the head of $r$, and all literals in the
	body of $r$ are true w.r.t.\ $M$.
	If $M$ is a stable model of a program $\Pi$, then all atoms in $M$ are supported.
	
	\begin{algorithm}
		\SetKwInOut{Input}{Input}
		\SetKwInOut{Output}{Output}
		\Input{A ground program $\Pi$}
		\Output{\coherent if $\mathit{SM}(\Pi) \neq \emptyset$. Otherwise, \inco}
		\Begin{
			$I := \emptyset$\;
			
			($\Pi$, $I$):= \textit{SimplifyProgram}($\Pi$, $I$); \hfill \tcp{remove redundant rules and atoms from $\Pi$}\label{ln:alg:simplify}
			$I$ := \textit{Propagate}($I$); \hfill \tcp{propagate deterministic consequences of $I$} \label{ln:alg:propagate}
			\uIf{$I$ is inconsistent \label{ln:alg:inconsistency}}
			{
				$r$ := \textit{CreateConstraint}($I$); \hfill \tcp{learning} \label{ln:alg:learning}
				\lIf{$B(r)$ = $\emptyset$}{
					\Return \inco; \hfill \tcp{$\Pi$ does not admit stable models}
				}
				
				$\Pi$ := $\Pi$ $\cup$ $\{r\}$\label{ln:alg:addconstraint}\;
				$I$ := \textit{RestoreConsistency}($I$, $\Pi$); \hfill \tcp{unroll until $I$ is consistent} \label{ln:alg:unroll}
			}
			\uElseIf{$I$ is total}
			{
				\lIf{\textit{CheckConsistency}($I$) \label{ln:alg:checkconsistency}} {\Return \coherent; \hfill \tcp{$I$ is a stable model}}
				$R$ := \textit{CreateConstraints}($I$); \hfill \tcp{create constraints for the failure} \label{ln:alg:createconstraintscheck}
				$\Pi$ := $\Pi$ $\cup$ $R$\label{ln:alg:addconstraintcheck}\;
				$I$ := \textit{RestoreConsistency}($I$, $\Pi$); \hfill \tcp{unroll until $I$ is consistent} \label{ln:alg:unrollcheck}			
			}
			\Else		
			{
				$I$ := \textit{RestartIfNeeded}($I$); \hfill \tcp{restart of the computation} \label{ln:alg:restart}
				$\Pi$ := \textit{DeleteConstraintsIfNeeded}($\Pi$); \hfill \tcp{deletion of learned constraints}
				$I$ := $I$ $\cup$ \textit{ChooseLiteral}($I$); \hfill \tcp{heuristic choice} \label{ln:alg:choice}
			}
			\textbf{goto}~\ref{ln:alg:propagate}\;
		}
		\caption{ComputeStableModel}\label{alg:mg}
	\end{algorithm}	
	\begin{function}
		$I$ := \textit{EagerPropagation}($I$)\label{ln:alg:begin}\;
		
		\If{$I$ is consistent} {
			$I'$ := \textit{PostPropagation}($I$)\; \label{ln:alg:post}
			\lIf{$I'$ = $\emptyset$}{\Return $I$\;}
			$I:=I \cup I'$\;		
		}
		\lIf{$I$ is inconsistent} {\Return $I$\;}
		\textbf{goto} \ref{ln:alg:begin}\;
		\caption{Propagate($I$)}\label{fn:propagate}
	\end{function}
	
	\subsection{CDCL Algorithm for Stable Model Computation} \label{sec:cdcl}
	The computation of a stable model is usually carried out by employing the Conflict-Driven Clause Learning (CDCL) algorithm~\cite{DBLP:journals/tc/Marques-SilvaS99,DBLP:conf/iccad/ZhangMMM01} with extensions specific to ASP \cite{DBLP:journals/aim/KaufmannLPS16}, reported here as Algorithm~\ref{alg:mg}.
	The algorithm takes as input a propositional program $\Pi$, and produces as output either \coherent, if $\Pi$ admits stable models, or \inco otherwise.
	
	The computation starts by applying polynomial simplifications to strengthen and/or remove redundant rules on the lines of methods employed in~\cite{DBLP:conf/ecai/GebserKNS08} and inspired by \cite{DBLP:conf/sat/EenS03} (see \cite{CarmineDodaro2015} for more details). 
	After the simplifications step, the non-chronological backtracking search starts.
	First, a partial interpretation $I$, initially empty, is extended with all the literals that can be deterministically inferred by applying some inference rule (propagation step, line~\ref{ln:alg:propagate}).
	Three cases are possible after a propagation step is completed:
	\begin{itemize}
	\item[$(i)$] $I$ is consistent but not total. In that case, an undefined literal $\ell$ (called branching literal) is chosen according to some heuristic criterion (line~\ref{ln:alg:choice}), and is added to $I$. Subsequently, a propagation step is performed that infers the consequences of this choice.
	\item[$(ii)$] $I$ is inconsistent, thus there is a conflict, and $I$ is analyzed. 
	The reason of the conflict is modeled by a fresh constraint $r$ (learning, line~\ref{ln:alg:learning}), computed in such a way to avoid the same conflict in the future search, e.g. using the \emph{first Unique Implication Point (UIP)} learning schema~\cite{DBLP:conf/iccad/ZhangMMM01}.
	If the learning procedure determines that the conflict cannot be avoided, i.e. the input program is incoherent, then the algorithm terminates returning $\inco$.
	Otherwise, the algorithm backtracks (i.e. choices and their consequences are undone, this is often referred to as \textit{unroll} in the following) until the consistency of $I$ is restored (line~\ref{ln:alg:unroll}) and $r$ is added to $\Pi$.
	\item[$(iii)$]x $I$ is total, the algorithm performs a consistency check on the interpretation $I$ (line~\ref{ln:alg:checkconsistency}).
	If $I$ is inconsistent the conflict is analyzed and a set of constraints is added to $\Pi$ (line~\ref{ln:alg:addconstraintcheck}). Otherwise, the algorithm terminates returning $I$.
	This check is required whenever the specific implementation of the CDCL algorithm lazily postpones some propagation inference which is required to assure the consistency of $I$.
	\end{itemize}
	
	For the performance of this search procedure, several details are crucial: learning effective constraints from inconsistencies, an effective propagation function as well as heuristics for restarting, constraint deletion, and for choosing literals.
	
	\paragraph{Propagation.}
	One of the key features of Algorithm~\ref{alg:mg} is the function~\ref{fn:propagate}, whose role is to extend the partial interpretation with the literals that can be deterministically inferred.
	In particular, a set of propagators is usually applied according to some priority sequence.
	Higher priority propagators are applied first (function $\mathit{EagerPropagation}$), while lower priority propagators are applied later (function $\mathit{PostPropagation}$).
	In the following, higher and lower priority propagators are referred to as \textit{eager} and \textit{post} propagators, respectively.
	An example of eager propagator is the \textit{unit propagator}.
	That is, given a partial interpretation $I$ consisting of literals, and a set of rules $\Pi$, unit propagation infers a literal $\ell$ to be true if there is a rule $r \in \Pi$ such that $r$ can be satisfied only by $I \cup \{\ell\}$.
	Consider a rule $r$ of the form (\ref{eq:rule}), its nogood representation is $C(r) = \{ \naf a_1, \ldots, \naf a_n, b_1, \ldots, b_j, \naf b_{j+1}, \ldots, \naf b_m \}$, which intuitively represents a constraint that is satisfied w.r.t. $I$ if and only if $r$ is satisfied w.r.t. $I$. Therefore, the negation of a literal $\ell \in C(r)$ is unit propagated w.r.t.\ $I$ and rule $r$ iff $C(r) \setminus \{ \ell \} \subseteq I$, since this represents the only way to satisfy the rule.
	The choice of the propagators used for the computation of stable models depends on the implementation of CDCL, which may vary the used propagators according to specific solving strategies or to features of the input program.
	As an example, to ensure that models are supported in presence of disjunctive rules with more than two atoms in the head, \clasp applies a (component-wise) \textit{shift} technique combined with unit propagation on the Clark completion of $\Pi$~\cite{DBLP:conf/ijcai/GebserKS13},  whereas \wasp employs a dedicated \emph{support propagator} as described in~\cite{DBLP:conf/ijcai/AlvianoD16}.
	
	\paragraph{Heuristic Choice.}
	The heuristic criteria used for selecting the branching literal play a crucial role in the CDCL algorithm.
	During the recent years, several heuristic strategies have been proposed.
	Among them, VSIDS~\cite{DBLP:conf/dac/MoskewiczMZZM01} has been shown to be successful in solving a large number of problems.
	The implementation of the state-of-the-art SAT solver \minisat~\cite{DBLP:conf/sat/EenS03} provides a further boost in the usage of VSIDS-like heuristics.
	Nowadays, ASP solvers \clasp and \wasp also use variants of the heuristic strategy proposed by \minisat.
	This strategy keeps an \textit{activity} value, initially set to 0, for each atom in $\mathit{At}(\Pi)$.
	When a literal $\ell=a$ or $\ell=\naf a$ is used for computing a learned constraint (e.g. during the computation of first UIP), the activity of $a$ is incremented by a value $inc$.
	The value of $inc$ is not static, instead it is multiplied by a constant slightly greater than 1 whenever a learned constraint is added to $\Pi$.
	Intuitively, this is done to give more importance to atoms which have been included in the recent learned constraints.
	The next branching literal is $\naf a$, where $a$ is the undefined atom with the highest value of activity (ties are broken randomly).
	In the following, the default heuristic coupled with the CDCL algorithm is assumed to be the \minisat heuristic. 
	
	\section{The External Interface of \wasp} \label{sec:interface}
	In this section, we describe the external interface of \wasp~\cite{DBLP:conf/lpnmr/AlvianoDLR15} for adding new propagators and heuristics.
	The architecture of \wasp allows by design the interaction of its internal CDCL algorithm (as described in Section~\ref{sec:cdcl}) with external algorithms developed according to the interface of methods described in Sections~\ref{sec:propagation} and~\ref{sec:heuristic}.
	During the computation of a stable model, and in particular when specific points of the computations are reached, \wasp performs calls to the corresponding methods of the external interface.
	Therefore, propagators and heuristics are algorithms providing an implementation of specific methods of the interface.
	In Sections~\ref{sec:propagation} and~\ref{sec:heuristic}, we describe such methods providing their \textit{contract}, i.e. the input parameter (Parameter), the output of the method (Return), the point of CDCL when the method is called (When), the conditions that must be true when the method is called (Preconditions), and the conditions that will be true when the method has completed its task (Postconditions).
	In order to simplify the presentation some technical details are omitted and the general description of the interface is not committed to a specific language.
	Moreover, we assume that \wasp takes as input a propositional program $\Pi$ and creates an interpretation $I$ initially set to $\emptyset$.
		
	\subsection{Propagators}\label{sec:propagation}
	The methods of the interface to add new propagators in \wasp are reported in the following and described in what follows in separate paragraphs.
	
	\paragraph{Method AttachLiterals.} This method associates a set of literals to the specific propagator. The contract of the method is the following:
	\begin{itemize}
		\item[] \textbf{Parameter:} none.
		\item[] \textbf{Return:} a set of literals $\L$.
		\item[] \textbf{When:} the method is called before the initial simplifications, i.e. before line \ref{ln:alg:simplify} of Algorithm~\ref{alg:mg}.
		\item[] \textbf{Preconditions:} the parsing of $\Pi$ is executed and the initial simplifications are not performed.
		\item[] \textbf{Postconditions:} literals in $\L$ are associated to the propagator.		
	\end{itemize}
	Literals in $\L$ are interpreted by \wasp as \textit{attached} to the propagator. That is, whenever a literal in $\L$ is added to or removed from the partial interpretation $I$, a notification is sent to the propagator using the methods described in the following (see \textit{OnLiteralTrue} and \textit{OnLiteralsUndefined}, respectively). Otherwise, literals which are not included in $\L$ are ignored.
	Intuitively, this method is used to limit the notifications only to a subset of literals of interest for the propagator.
	
	\paragraph{Method Simplify.} This method can be used to further simplify the input program. The contract of the method is the following:
	\begin{itemize}
		\item[] \textbf{Parameter:} none.
		\item[] \textbf{Return:} a set of literals.
		\item[] \textbf{When:} the method is called during the execution of $\mathit{SimplifyProgram}$ (line \ref{ln:alg:simplify} of Algorithm~\ref{alg:mg}), i.e. after all simplifications implemented by \wasp.
		\item[] \textbf{Preconditions:} initial simplifications of the input program have been performed and $I$ is consistent.
		\item[]	\textbf{Postconditions:} literals returned by the method are added to $I$.
	\end{itemize}
	During the simplifications, a custom propagator may identify a set of literals that must be included in $I$. Stated differently, the propagator can return a set of literals that will be always included in all stable models of $\Pi$.

	\paragraph{Method OnLiteralTrue.} This method can be used to implement eager propagators. The contract of the method is the following:
	\begin{itemize}
		\item[] \textbf{Parameter:} a literal $\ell \in \L$ that has been added to $I$.
		\item[] \textbf{Return:} a set of literals.
		\item[] \textbf{When:} the method is called during the execution of \textit{EagerPropagation} (line~\ref{ln:alg:begin} of \ref{fn:propagate}).
		\item[] \textbf{Preconditions:} the interpretation $I$ is consistent, $\ell \in (\L \cap I)$.
		\item[] \textbf{Postconditions:} literals returned by the method are added to $I$.		
	\end{itemize}
	A literal $\ell$ is added to the interpretation $I$. Such a literal may lead to the inference of other literals, which are returned as output by this method and that will be later on added to the interpretation $I$ by \wasp.
	
	\paragraph{Method OnLiteralsTrue.} This method can be used to implement post propagators. The contract of the method is the following:
	\begin{itemize}
		\item[] \textbf{Parameter:} a set of literals $L \subseteq \L$.
		\item[] \textbf{Return:} a set of literals.
		\item[] \textbf{When:} the method is called during the execution of $\mathit{PostPropagation}$ (line~\ref{ln:alg:post} of \ref{fn:propagate}).
		\item[] \textbf{Preconditions:} the interpretation $I$ is consistent, $L \subseteq (\L \cap I)$.
		\item[] \textbf{Postconditions:} literals returned by the method are added to $I$.		
	\end{itemize}
	The parameter $L$ includes the latest heuristic choice and literals added to $I$ during the execution of $\mathit{EagerPropagation}$. As the previous method, it returns a set of literals which will be later on added to the interpretation $I$ by \wasp.
	
	\paragraph{Method OnLiteralsUndefined.} The method can be used by the propagator to keep track of modifications of $I$. The contract of the method is the following:
	\begin{itemize}
		\item[] \textbf{Parameter:} a set of literals $L \subseteq \L$.
		\item[] \textbf{Return:} none.
		\item[] \textbf{When:} this method is called either during the execution of methods $\mathit{RestoreConsistency}$ or $\mathit{RestartIfNeeded}$ (lines~\ref{ln:alg:unroll} and~\ref{ln:alg:restart} of Algorithm~\ref{alg:mg}, respectively).
		\item[] \textbf{Preconditions:} an unroll of the interpretation $I$ has been performed, $I$ is consistent, $L \subseteq \L$, $(L \cup \overline{L}) \cap I = \emptyset$.
		\item[] \textbf{Postconditions:} none.
	\end{itemize}
	Literals in $L$ have been removed from the partial interpretation $I$ by \wasp, thus they are undefined, e.g. after a conflict or a restart.

	\paragraph{Method GetReasonForLiteral.} The method can be used for providing an explanation for the inference of a literal. The contract of the method is the following:
	\begin{itemize}
		\item[] \textbf{Parameter:} a literal $\ell$.
		\item[] \textbf{Return:} a constraint $r$.
		\item[] \textbf{When:} this method is called for each literal $\ell$ added to the interpretation by methods \textit{OnLiteralTrue} and \textit{OnLiteralsTrue}.
		\item[] \textbf{Preconditions:} $\ell$ is the literal (resp. one of the literals) returned by the method \textit{OnLiteralTrue} (resp. \textit{OnLiteralsTrue}).
		\item[] \textbf{Postconditions:} let $r$ be the constraint returned by the method, the negation of $\ell$ is in $r$ and all literals in $r$ but $\ell$ are true w.r.t. the partial interpretation $I$, $r$ is added to $\Pi$.
	\end{itemize}
	Whenever a literal $\ell$ is inferred by one of the methods \textit{OnLiteralTrue} and \textit{OnLiteralsTrue}, an explanation of the inference is needed to proper implement learning techniques.
	Such an explanation is modeled by a constraint $r$. More formally, let $S=\{\ell_1, \ldots, \ell_m\}$ be a set of literals such that if $S \subseteq I$ then $\ell \in I_1$ for each stable model $I_1 \supseteq I$ of $\Pi$, the constraint $r$ is of the form $\leftarrow \ell_1, \ldots, \ell_m, \naf \ell$. 

	\paragraph{Method CheckStableModel.} The method can be used to check the consistency of $I$ with respect to the propagator. The contract of the method is the following:
	\begin{itemize}
		\item[] \textbf{Parameter:} a set of literals representing the interpretation $I$.
		\item[] \textbf{Return:} a Boolean value.
		\item[] \textbf{When:} this method is called during the execution of $\mathit{CheckConsistency}$ (line~\ref{ln:alg:checkconsistency} of Algorithm~\ref{alg:mg}).
		\item[] \textbf{Preconditions:} $I$ is total and consistent w.r.t. $\Pi$.
		\item[] \textbf{Postconditions:} If the method returns \textit{true}, \wasp terminates its execution. If the method returns \textit{false}, the method \textit{GetReasonsForCheckFailure} is called.		
	\end{itemize}
	The propagator may lazily postpone some inference that is required to assure the consistency of $I$, therefore this method returns \textit{true} if $I$ is consistent with respect to the propagator, and \textit{false} otherwise.
	
	\paragraph{Method GetReasonsForCheckFailure.} The method can be used for providing an explanation of the stable model check failure. The contract of the method is the following:
	\begin{itemize}
		\item[] \textbf{Parameter:} none.
		\item[] \textbf{Return:} a set of constraints.
		\item[] \textbf{When:} this method is called whenever the method $\mathit{CheckStableModel}$ returns \textit{false}.
		\item[] \textbf{Preconditions:} method \textit{CheckStableModel} returned \textit{false}, $I$ is total and consistent w.r.t. $\Pi$.
		\item[] \textbf{Postconditions:} let $R$ be the set of constraints returned by the method, for each constraint $r \in R$ all literals in $r$ are true w.r.t. $I$, $R$ is added to $\Pi$ and the computation of \wasp restarts.
	\end{itemize}
	The failure of \textit{CheckStableModel} is explained in terms of constraints that are implicitly violated by the interpretation $I$. Indeed, the consistency check fails when the interpretation is inconsistent with respect to the propagator. Constraints in $R$ are later on added to $\Pi$ by \wasp (line~\ref{ln:alg:addconstraintcheck} of Algorithm~\ref{alg:mg}).
	
	\subsection{Heuristic}\label{sec:heuristic}
	The interface of \wasp also allows the definition of custom heuristics that modify the default \minisat heuristic, using the methods described in the following.
	Methods included in the interface were chosen by looking at modern general-purpose and domain-specific heuristics.
	Note that an algorithm modifying the heuristic can also use the methods described in the previous section.
	
	\paragraph{Method OnConflict.} The method can be used to keep track that a conflict occurred during the search. The contract of the method is the following:
	\begin{itemize}
		\item[] \textbf{Parameter:} none.
		\item[] \textbf{Return:} none.
		\item[] \textbf{When:} this method is called whenever the partial interpretation $I$ is inconsistent (i.e. line~\ref{ln:alg:inconsistency} of Algorithm~\ref{alg:mg}).
		\item[] \textbf{Preconditions:} the partial interpretation $I$ is inconsistent, i.e. a conflict occurred during the search.
		\item[] \textbf{Postconditions:} none.
	\end{itemize}
	The number of conflicts occurring during the search is a parameter that is often used by many look-back heuristics, as the \minisat one. Therefore, this method provides a convenient way to keep track of the conflicts occurred.
		
	\paragraph{Method OnLitInConflict.} The method can be used to keep track of literals that are used during the computation of the first UIP. The contract of the method is the following:
	\begin{itemize}
		\item[] \textbf{Parameter:} a literal $\ell$.
		\item[] \textbf{Return:} none.
		\item[] \textbf{When:} this method is called during the execution of $\mathit{CreateConstraint}$ and $\mathit{CreateConstraints}$ (lines \ref{ln:alg:learning} and \ref{ln:alg:createconstraintscheck} of Algorithm~\ref{alg:mg}, respectively).
		\item[] \textbf{Preconditions:} a learned constraint $r$ is created after a conflict and $\ell$ is a literal used during the computation of first UIP.
		\item[] \textbf{Postconditions:} none.
	\end{itemize}
	Literals used during the computation of the first UIP are usually considered good candidates as branching literals by look-back techniques \cite{DBLP:conf/sat/EenS03}. Therefore, this method provides a convenient way to keep track of such literals.
	
	\paragraph{Method OnLearningConstraint.} The method can be used to keep track of constraints learned during the search. The contract of the method is the following:
	\begin{itemize}
		\item[] \textbf{Parameter:} a constraint $r$.
		\item[] \textbf{Return:} none.
		\item[]	\textbf{When:} this method is called whenever a new constraint $r$ is added to the input program (lines \ref{ln:alg:addconstraint} and \ref{ln:alg:addconstraintcheck} of Algorithm~\ref{alg:mg}). 
		\item[] \textbf{Preconditions:} a learned constraint $r$ is added to $\Pi$.
		\item[] \textbf{Postconditions:} none.
	\end{itemize}	
	
	\paragraph{Method OnRestart.} The method can be used to keep track of restarts occurring during the computation. The contract of the method is the following:
	\begin{itemize}
		\item[] \textbf{Parameter:} none.
		\item[] \textbf{Return:} none.
		\item[] \textbf{When:} this method is called during the execution of $\mathit{RestartsIfNecessary}$ (line \ref{ln:alg:restart} of Algorithm~\ref{alg:mg}).
		\item[] \textbf{Preconditions:} the solver restarted the computations from scratch and no branching choices are made.
		\item[] \textbf{Postconditions:} none.
	\end{itemize}
	
	\paragraph{Method Init-\minisat.} The method can be used to initialize the \minisat activities of some atoms. The contract of the method is the following:
	\begin{itemize}
		\item[] \textbf{Parameter:} none.
		\item[] \textbf{Return:} a set of elements of the form $(a,v)$, where $a\in At(\Pi)$ and $v \geq 0$ is an integer.
		\item[] \textbf{When:} this method is called by \wasp after the initial simplification and before the first propagation, i.e. after line \ref{ln:alg:simplify} of Algorithm~\ref{alg:mg}.
		\item[] \textbf{Preconditions:} the simplifications are performed, no branching choices are made and the \minisat activities of all atoms in the program are set to 0.
		\item[] \textbf{Postconditions:} let $S$ be the set of elements of the form $(a,v)$ returned by the method where $a \in At(\Pi)$ and $v \geq 0$, the \minisat activity of the atom $a$ is set to $v$ for each element $(a,v)$ in $S$.
	\end{itemize}
	The default implementation of \minisat heuristic initializes the activity of all atoms to 0. Therefore, this method can be used to initialize the activity of atoms to a different value. Intuitively, since the \minisat heuristic selects atoms with the highest activity value, a proper selection of initial activity values can be useful to influence the first choices made by the solver.
	
	\paragraph{Method Factor-\minisat.}	The method can be used to associate an amplifying factor to the \minisat activities of some atoms. The contract of the method is the following:
	\begin{itemize}
		\item[] \textbf{Parameter:} none.
		\item[] \textbf{Return:} a set of elements of the form $(a,v)$, where $a\in At(\Pi)$ and $v \geq 0$ is an integer.
		\item[] \textbf{When:} this method is called by \wasp after the initial simplification and before the first propagation, i.e. after line \ref{ln:alg:simplify} of Algorithm~\ref{alg:mg}.
		\item[] \textbf{Preconditions:} the simplifications are performed, no branching choices are made and the \minisat activities of all atoms are initialized.
		\item[] \textbf{Postconditions:} let $S$ be the set of elements of the form $(a,v)$ returned by the method where $a \in At(\Pi)$ and $v \geq 0$, the \minisat activity of the atom $a$ is associated to the amplifying factor $v$ for each element $(a,v)$ in $S$.
	\end{itemize}
	The amplifying factor will be used by \wasp during the computation of the heuristic choice, in particular the activity value of an atom is multiplied by its amplifying factor.
	
	\paragraph{Method Sign-\minisat.} The method can be used to provide a priority on the sign of literals, that an atom will be selected with a specific polarity. The contract of the method is the following:
	\begin{itemize}
		\item[] \textbf{Parameter:} none.
		\item[] \textbf{Return:} a set of elements of the form $(a,v)$, where $a\in At(\Pi)$ and $v\in\{pos, neg\}$.
		\item[] \textbf{When:} this method is called by \wasp after the initial simplification and before the first propagation, i.e. after line \ref{ln:alg:simplify} of Algorithm~\ref{alg:mg}.
		\item[] \textbf{Preconditions:} the simplifications are performed, no branching choices are made and the \minisat activities of all atoms are initialized.
		\item[] \textbf{Postconditions:} let $S$ be the set of elements of the form $(a,v)$ returned by the method where $a \in At(\Pi)$ and $v \in \{pos, neg\}$, the atom $a$ is associated with the polarity $v$.
	\end{itemize}
	The sign of atoms allows to specify a preference of the choice polarity of atoms. In particular, let $a$ be an atom and $v \in \{pos, neg\}$ its associated polarity, whenever the atom $a$ is selected as branching atom by the \minisat heuristic, the corresponding branching literal is $\naf a$ if $v=neg$, otherwise it is $a$. 
	
	\paragraph{Method SelectLiteral.} The method can be used to replace the default \minisat heuristic or to trigger a restart of the computation or an unroll.
	The contract of the method is the following:
	\begin{itemize}
		\item[] \textbf{Parameter:} none.
		\item[] \textbf{Return:} an element of the following form: \textit{(i)} (\textit{choice}, $\ell$), where $\ell$ is an undefined literal w.r.t. $I$ \textit{(ii)} (\minisat, $n$), where $n \geq 0$ \textit{(iii)} (\textit{unroll}, $\ell$) where $\ell$ is a true or false literal w.r.t. $I$ or \textit{(iv)} (\textit{restart}).
		\item[] \textbf{When:} this method is called by \wasp whenever an heuristic choice is required, i.e. during the execution of \textit{ChooseLiteral} (line~\ref{ln:alg:choice} of Algorithm~\ref{alg:mg}).
		\item[] \textbf{Preconditions:} the interpretation $I$ is consistent, 
		\item[] \textbf{Postconditions:} in case \textit{(i)} the undefined literal $\ell$ is selected as branching literal and added to $I$; in case \textit{(ii)} the custom heuristic is disabled for the subsequent $n$ choices switching to the \minisat one, or the \minisat heuristic is enabled permanently in case $n$=0; in case \textit{(iii)} all the choices are retracted until $\ell$ is not included anymore in $I$; in case \textit{(iv)} \wasp performs a complete restart of the computation.	
	\end{itemize}

	\subsection{Implementation}
	The interface described in the previous section has been implemented as an extension of \wasp.
	Current implementation supports \textit{perl} and \textit{python} scripts for obtaining fast prototypes, and \textit{C++} in case better performance is needed.
	It is important to emphasize that \textit{C++} implementations must be integrated in the \wasp binary at compile time, whereas \textit{perl} and \textit{python} scripts are specified by means of files given as parameters for \wasp, thus they do not require changes and recompilation of its source code.
	In particular, \textit{perl} or \textit{python} propagators and heuristics are files that are provided to \wasp by means of a command line option.
	Note that such files do not contain any reference to the internal data structures of \wasp.
	The communication between the solver and the file is done by means of a message passing protocol, which is however transparent to the developer of propagators and heuristics.
	The source code and the documentation are available at \url{alviano.github.io/wasp/}.
	
	\begin{figure}
		\figrule
		$$
		\begin{array}{lrl}	
		\multicolumn{3}{l}{\% \ \mathit{Guess \ matches}}\\	
		r_1: & \mathit{match(M,W)} \quad \leftarrow 	 &\mathit{man(M)}, \ \mathit{woman(W)}, \ \naf \mathit{nmatch(M,W)}\\
		r_2: &\mathit{nmatch(M,W)} \quad \leftarrow & \mathit{man(M)}, \ \mathit{woman(W)}, \ \naf \mathit{match(M,W)}\\
		\\
		\multicolumn{3}{l}{\% \ \mathit{No \ polygamy}}\\
		r_3: &\leftarrow & \mathit{match(M_1,W)},\ \mathit{match(M_2,W)},\ \mathit{M_1} \neq \mathit{M_2}\\
		r_4: &\leftarrow & \mathit{match(M,W_1)},\ \mathit{match(M,W_2)},\ \mathit{W_1} \neq \mathit{W_2}\\
		\\
		\multicolumn{3}{l}{\% \ \mathit{No \ singles}}\\
		r_5: &\mathit{married(M)} \quad \leftarrow & \mathit{match(M,W)} \\
		r_6: &\leftarrow & \mathit{man(M)}, \ \naf \mathit{married(M)} \\
		\\
		\multicolumn{3}{l}{\% \ \mathit{Strong \ stability\ condition}}\\	
		r_7: &\leftarrow & \mathit{match(M,W_1)}, \ \mathit{match(M_1,W)}, \ \mathit{W_1} \neq \mathit{W},\\
		& & \mathit{pref(M,W_1,SM_1)},\ \mathit{pref(M,W,SM)},\ \mathit{SM} >  \mathit{SM_1},\\
		& & \mathit{pref(W,M_1,SW_1)},\ \mathit{pref(W,M,SW)},\ \mathit{SW} \geq \mathit{SW_1}\\
		
		\end{array}
		$$
		\caption{ASP Encoding of the Stable Marriage Problem.}\label{fig:encoding:stable}
		\figrule
	\end{figure}
		
	\section{Examples of Usage}
	In this section, we provide two usage examples of the interface described in the previous section.
	Section~\ref{sec:example:stablemarriage} reports the implementation of two propagators for solving the \textit{Stable Marriage} problem.
	Section~\ref{sec:example:casp} shows a proof-of-concept implementation of a CASP solver.
	Section~\ref{sec:example:vsids} describes the instantiation of the interface for implementing the VSIDS heuristic as proposed by the solver \textsc{chaff}~\cite{DBLP:conf/dac/MoskewiczMZZM01}.
	
	\begin{algorithm}[t]
		\textbf{Global variables:} $\Pi$, $C$\;
		\BlankLine\BlankLine		
		\textbf{Method: } CheckStableModel($I$)\\
		\Indp
		$C$ := $\emptyset$\;
		$\mathit{Matches}$ := $\{a \mid a \in \mathit{At}(\Pi) \cap I, \ a \mathit{\ is \ of \ the\ form\ } \mathit{match}(m,w)\}$\;
		
		\For{$match(m,w_1) \in \mathit{Matches}$}
		{
			\For{$match(m_1,w) \in \mathit{Matches}$ \textbf{\textup{and}} $m \neq m_1$}
			{
				\tcp{$\mathit{pref}(m,w_1,sm_1)$, $\mathit{pref}(m,w,sm)$, $\mathit{pref}(w,m_1,sw_1)$, $\mathit{pref}(w,m,sw) \in \mathit{At}(\Pi)$}
				\lIf{$sm > sm_1$ \textbf{\textup{and}} $sw \geq sw_1$} {
					$C$ := $C \cup \{\leftarrow match(m,w_1), \ match(m_1,w)\}$\;
				}
			}			
		}
		\Return $C = \emptyset$; \hfill \tcp{no constraints to add: model is stable and return \textit{true}}
		\Indm		
		\BlankLine\BlankLine
		\textbf{Method: } GetReasonForCheckFailure()\\
		\Indp			
		\Return $C$; \hfill \tcp{$C$ is modified by CheckStableModel}
		\Indm
		\caption{LazyStableMarriage}\label{alg:stablemarriage:lazy}
	\end{algorithm}			
	\begin{algorithm}[t]
		\textbf{Global variables:} $\Pi$, $\ell_1$\;
		\BlankLine\BlankLine		
		\textbf{Method: } AttachLiterals()\\ 
		\Indp
		\Return $\{a \mid a \in \mathit{At}(\Pi), \ a \mathit{\ is \ of \ the\ form\ } \mathit{match}(m,w_1)\}$\;
		\Indm
		\BlankLine\BlankLine
		\textbf{Method: } OnLiteralTrue($\ell$)\\ 
		\Indp
		$\ell_1$ := $\ell$; \quad $L$ := $\emptyset$; \quad $\mathit{Matches}$ := $\{a \mid a \in \mathit{At}(\Pi), \ a \mathit{\ is \ of \ the\ form\ } \mathit{match}(m_1,w), m \neq m_1\}$\;
		
		\tcp{$\ell$ is of the form $\mathit{match}(m,w_1)$}
		\For{$match(m_1,w) \in \mathit{Matches}$}
		{
			\tcp{$\mathit{pref}(m,w_1,sm_1)$,$\mathit{pref}(m,w,sm)$,$\mathit{pref}(w,m_1,sw_1)$,$\mathit{pref}(w,m,sw)\in \mathit{At}(\Pi)$}
			\tcp{$\mathit{pref}(m_1,w_1,sm_2)$,$\mathit{pref}(m_1,w,sm_3)$,$\mathit{pref}(w_1,m,sw_2)$,$\mathit{pref}(w_1,m_1,sw_3)\in \mathit{At}(\Pi)$}
			\If{$sm > sm_1$ \textbf{\textup{and}} $sw \geq sw_1$ \textbf{\textup{or}} $sm_2 > sm_3$ \textbf{\textup{and}} $sw_3 \geq sw_2$} {
				$L$ := $L \cup \{\naf \mathit{match}(m_1,w)\}$\;
			}
		}
		\Return $L$\;
		\Indm
		\BlankLine\BlankLine
		\textbf{Method: } GetReasonForLiteral($\ell$)\\ 
		\Indp
		\tcp{let $\ell$ be of the form $\naf \mathit{match}(m_1,w)$}
		\Return $\leftarrow \mathit{match}(m_1,w), \ell_1$; \hfill \tcp{$\ell_1$ is modified by OnLiteralTrue}
		\Indm
		\caption{EagerStableMarriage}\label{alg:stablemarriage:eager}
	\end{algorithm}
	
	\subsection{Propagators for Stable Marriage}\label{sec:example:stablemarriage}
	In this section, the interface of \wasp is used for adding new external propagators for solving the \textit{Stable Marriage} problem.
	The role of propagators is to replace the instantiation of some problematic constraints as described by \citeN{DBLP:journals/tplp/CuteriDRS17}.
	The Stable Marriage problem can be described as follows: 
	given $n$ men and $m$ women, where each person has a preference order over the opposite sex, marry them so that the marriage is stable. 
	In this case, the marriage is  said to be stable if there is no pair $(m', w')$ for which both partners would rather be married with each other than their current partner.
	In particular, an ASP encoding of the Stable Marriage problem used for the fourth ASP Competition is reported in Figure~\ref{fig:encoding:stable}, where the two disjoint sets of men $M$ and women $W$ are encoded by instances of the predicates $\mathit{man}$ and $\mathit{woman}$, respectively.
	Instances of the predicate $\mathit{pref}$ represent preferences of men to women and of women to men. In particular, an atom of the form $\mathit{pref(m,w,n)}$ encodes the preference of man $m$ to woman $w$.
	
	The first two rules of the encoding define the search space by guessing a match of men and women, while subsequent constraints filter out matches that do not satisfy the requirements.
	As argued in \cite{DBLP:journals/tplp/CuteriDRS17}, the last constraint ($r_7$), which guarantees that the stability condition is not violated, might be problematic to be evaluated by classical solving strategies.
	Indeed, the instantiation of this constraint requires the generation of a huge number of ground constraints.
	Thus, a possible solution is to replace $r_7$ by means of external propagators.
	In particular, given the encoding reported in Figure~\ref{fig:encoding:stable}, the idea is to evaluate only rules $r_1$--$r_6$, while $r_7$ is removed from the encoding.
	Then, stable models of the resulting program violating $r_7$ are filtered out by means of ad-hoc implementations.
	Two strategies were proposed and analyzed in \cite{DBLP:journals/tplp/CuteriDRS17}.
	The first one was called \textit{lazy} and basically lazily instantiates $r_7$ whenever it is violated by a stable model candidate.
	The second strategy was called \textit{eager}. In this case, the idea is to simulate the unit propagation over $r_7$ during the computation of the stable model candidate.
	Both strategies can be implemented in \wasp using the interface described in Section~\ref{sec:interface}.
	
	\paragraph{Lazy.}
	Lazy approach is reported in Algorithm~\ref{alg:stablemarriage:lazy}. This strategy uses methods $\mathit{CheckStableModel}$ and $\mathit{GetReasonForCheckFailure}$.
	Whenever a stable model candidate is found, \wasp calls the method $\mathit{CheckStableModel}$, which checks whether the stability condition modeled by constraint $r_7$ is violated.
	If there is no violation then the algorithm terminates returning $\mathit{true}$, thus witnessing the stability of the model candidate.
	Otherwise, it produces a set of violated constraints which are later on added to \wasp using the method $\mathit{GetReasonForCheckFailure}$.
	
	\paragraph{Eager.}
	Eager approach is reported in Algorithm~\ref{alg:stablemarriage:eager}. 
	In contrast to the previous strategy that aims at adding violated constraints when a stable model candidate is found, this one evaluates the constraints during the computation of the stable model.
	Thus, constraint $r_7$ is never instantiated in practice but its inference is simulated by an ad-hoc procedure implemented for that purpose.
	In particular, this strategy takes advantage of methods $\mathit{AttachLiterals}$, $\mathit{OnLiteralTrue}$ and $\mathit{GetReasonForLiteral}$.
	Method $\mathit{AttachLiterals}$ returns a list of all literals of the form $\mathit{match}(m,w)$, thus whenever an atom of this kind is propagated as true the method $\mathit{OnLiteralTrue}$ is called.
	The role of $\mathit{OnLiteralTrue}$ is to simulate unit propagation inferences over constraint $r_7$.	

\begin{algorithm}[t]
		\textbf{Global variables:} $\Pi$, $C$, $R$\;
		\BlankLine\BlankLine		
		\textbf{Method: } AttachLiterals()\\ 
		\Indp
		$C$ := $\{a \mid a \in \mathit{At}(\Pi), \ a \mathit{\ is \ of \ the\ form\ } \mathit{required}(c)\}$\;
		$C$ := $C \cup \{a \mid a \in \mathit{At}(\Pi), \ a \mathit{\ is \ of \ the\ form\ } \mathit{cspdomain}(d)\}$\;	
		$C$ := $C \cup \{a \mid a \in \mathit{At}(\Pi), \ a \mathit{\ is \ of \ the\ form\ } \mathit{cspvar}(x,n,m)\}$\;				
		\Return $\emptyset$\;
		\Indm
		\BlankLine\BlankLine		
		\textbf{Method: } CheckStableModel($I$)\\
		\Indp
		$\mathit{T}$ := $\{a \mid a \in C \cap I\}$; \hfill \tcp{identify true irregular and special atoms}

		$\mathcal{C}$ := $\mathit{Atoms2Constraints(T)}$; \hfill \tcp{implements $\gamma$}
		
		$(\mathit{res}, S)$ := $\mathit{ConstraintSolver} ( \mathcal{C} )$; \hfill \tcp{call an external solver}
		\eIf{$\mathit{res}$} {
			$\mathit{Print}(S)$;  \hfill \tcp{print the solution}
		}
		{
			$R$ := $T$; \hfill \tcp{trivial reason of failure}
		}
		
		\Return $\mathit{res}$; \hfill \tcp{no constraints to add: model is stable and return \textit{true}}
		\Indm		
		\BlankLine\BlankLine
		\textbf{Method: } GetReasonForCheckFailure()\\
		\Indp			
		\Return $\leftarrow l_1, \ldots, l_n$; \hfill \tcp{$R=\{l_1, \dots, l_n\}$ is modified by CheckStableModel}
		\Indm
		\caption{A naive CASP solver}\label{alg:casp}
	\end{algorithm}
\subsection{Implementation of a CASP solver}\label{sec:example:casp}

In this section, we sketch how the interface of \wasp can be used for implementing a naive CASP solver.
Our example is based on a fragment of the EZ language, as described in~\cite{DBLP:journals/tplp/BalducciniL17,DBLP:conf/iclp/SusmanL16}, and is inspired by the solving approach of \textsc{clingcon}~\cite{DBLP:journals/tplp/OstrowskiS12}. 

An EZ program $P$ is a triple $(\Pi,\mathcal{B}, \gamma)$, where $\Pi$ is ASP program, $\mathcal{B}$ is a set of \textit{constraints} of a CSP (e.g., linear constraints)~\cite{DBLP:reference/fai/2} and $\gamma$ is an injective function from the set of irregular atoms $C \subseteq \mathit{atoms}(\Pi)$ to $\mathcal{B}$. An answer set of $P$ is an answer set $X$ of $\Pi$ such that $\bigcup_{p \in X \cap C} \{\gamma(p)\}$ has a solution~\cite{DBLP:conf/iclp/SusmanL16}.

We implement the semantics by using the propagator reported in Algorithm~\ref{alg:casp}, which basically checks whether the CSP associated to the current candidate answer set has solution. If this is the case, the solution is printed, otherwise a reason for the failure of the check is  a single constraint having in the body all irregular atoms that are true (see function GetReasonForCheckFailure).
For completeness, the role of function AttachLiterals in Algorithm~\ref{alg:casp} is simply to intercept irregular atoms. In our example, all irregular atoms are expected to be of the form $\mathit{required(\cdot)}$, and the specification of domain variables use the special atoms $\mathit{cspdomain}(\cdot)$ and $\mathit{cspvar}(\cdot)$. This format is compliant with the output of the preprocessor of \textsc{ezsmt}~\cite{DBLP:conf/iclp/SusmanL16}. 
We refer the reader to~\cite{DBLP:journals/tplp/BalducciniL17,DBLP:conf/iclp/SusmanL16} for more details.

We remark that the goal of this example is to show the applicability of the interface. Therefore, the simple CASP solving strategy presented here is not expected to be efficient, since it does not include all sophisticated techniques proposed by state-of-the-art CASP solvers that would complicate the description.

		\begin{algorithm}
	\textbf{Global variables:} $\Pi$, $I := \emptyset$, $\mathit{scores} :=[]$, $\mathit{conflicts} := 0$;
	\BlankLine\BlankLine		
	\textbf{Method: } AttachLiterals()\\ 
	\Indp
	\lFor{$a \in \mathit{At}(\Pi)$} {
		$\mathit{scores[a]}$ := 0; \hfill \tcp{init scores of atoms}
	}
	\Return $\{a \mid a \in \mathit{At}(\Pi)\} \cup \{\naf a \mid a \in \mathit{At}(\Pi)\}$\;
	\Indm
	\BlankLine\BlankLine
	\textbf{Method: } OnLiteralsTrue($L$)\\ 
	\Indp
	$I$ := $I \ \cup \ L$\;
	\Return $\emptyset$; \hfill \tcp{No inferences done here}
	\Indm
	\BlankLine\BlankLine
	\textbf{Method: } OnUnrollLiterals($L$)\\ 
	\Indp
	$I$ := $I \ \setminus \ L$\;
	\Return \;
	\Indm
	\BlankLine\BlankLine
	\textbf{Method: } OnLearningConstraint($r$)\\ 
	\Indp
	\lFor{$a \in B(r)$}{
		$\mathit{scores[a]}$ := $\mathit{scores[a]}$ + 1; \hfill \tcp{atoms bumping}
	}		
	\Return \;
	\Indm
	\BlankLine\BlankLine	
	\textbf{Method: } OnConflict()\\ 
	\Indp		
	$\mathit{conflicts}$ := $\mathit{conflicts}$ + 1\;
	\If{$\mathit{conflicts} = 256$}{
		$\mathit{conflicts}$ := 0\;
		\lFor{$a \in \mathit{At}(\Pi)$}{
			$\mathit{scores[a]}$ := $\mathit{scores[a]} \div 2$; \hfill \tcp{atoms rescoring}
		}
		\textbf{sort}$\mathit{(scores, descending)}$; \hfill \tcp{sort scores in descending order}
	}
	\Return \;
	\Indm
	\BlankLine\BlankLine
	\textbf{Method: } SelectLiteral()\\ 
	\Indp
	$a$ := $\mathit{GetFirstUndefined}(\mathit{scores}, I)$; \hfill \tcp{select the first undefined atom}
	\Return $\naf a$\;
	\Indm
	\BlankLine\BlankLine
	\caption{VSIDS}\label{alg:heuristics:vsids}
\end{algorithm}	
	\subsection{Implementation of VSIDS Heuristic}\label{sec:example:vsids}
	In this section, the interface of \wasp is used for implementing the general-purpose variable state independent decaying sum (VSIDS) heuristic.
	VSIDS was proposed in the SAT solver \textsc{chaff}~\cite{DBLP:conf/dac/MoskewiczMZZM01} and it inspired several modern heuristics, including the \minisat heuristic.
	The basic idea is to store atoms in a list and to associate a numerical score to each atom, initially set to 0.
	After learning a constraint, the score of the atoms included in its body is incremented (called \textit{bumping}).
	The score of each atom is halved every 256 conflicts (called \textit{rescoring}) and the list of atoms is sorted descending according to their new score.
	Whenever a choice is required the first undefined atom in the list is selected.
	An implementation of the VSIDS heuristic as described in \cite{DBLP:conf/sat/BiereF15} using the interface of \wasp is reported in Algorithm~\ref{alg:heuristics:vsids}.
	Method $\mathit{AttachLiterals}$ is used to initialize the scores of atoms, while $\mathit{OnLiteralsTrue}$ and $\mathit{OnUnrollLiteral}$ are used to store and update the partial interpretation $I$.
	Method $\mathit{OnLearningConstraint}$ is used to bump the scores of atoms appearing in the constraint, while method $\mathit{OnConflict}$ is used to count the number of conflicts and to periodically update the scores of atoms.
	Finally, $\mathit{SelectLiteral}$ is used when a choice is needed.
	
	Note that modern implementations of VSIDS also bump the scores of atoms when they are used for the computation of a learned constraint.
	In this case, method $\mathit{OnLitInConflict}$ of the interface can be used.
		
	\section{Successful Applications}\label{sec:application}
	The interface of \wasp described in Section~\ref{sec:interface} has been already used in the literature to enhance the standard solving capabilities of \wasp~\cite{DBLP:journals/tplp/CuteriDRS17,DBLP:journals/tplp/DodaroGLMRS16,DBLP:conf/aiia/DodaroRS16}. In the following, such applications are reviewed discussing how the interface was used and its impact on the performance of \wasp.

	\subsection{Propagators}
	The interface of \wasp was used in \cite{DBLP:journals/tplp/CuteriDRS17,DBLP:conf/aiia/DodaroRS16} for avoiding the instantiation of some problematic constraints as shown in Section~\ref{sec:example:stablemarriage}.
	In particular, three propagators were proposed called \textit{lazy}, \textit{eager} and \textit{post}.
	The first two propagators use the same methods as described in Section~\ref{sec:example:stablemarriage}.
	In particular, methods $\mathit{CheckStableModel}$ and $\mathit{GetReasonForCheckFailure}$ were used for implementing the lazy propagator, while $\mathit{OnLiteralTrue}$ and $\mathit{GetReasonForLiteral}$ were used for implementing the eager propagator.
	In addition, the post propagator was implemented using $\mathit{OnLiteralsTrue}$ and $\mathit{GetReasonForLiteral}$.
	Those propagators were evaluated on three benchmarks, namely Stable Marriage, Packing and Natural Language Understanding. In the following, we review the main results obtained in \cite{DBLP:journals/tplp/CuteriDRS17}.
	
	\paragraph{Stable Marriage.}
	The experiment was executed on randomly generated instances.
	Basically, instances were generated as follows: each man (resp. woman) gives the same preference to each woman (resp. man), so that the stability condition is never violated.
	Then, a value percentage \textit{k} of preferences was also considered, i.e., each man (resp. woman) gives the same preference to all the women (resp. men) but to \textit{k}\% of them a lower preference is given.
	For each considered percentage \textit{k}, 10 randomly generated instances were considered.
	Results show that for instances where the value of \textit{k} is small (up to 50\%) the lazy approach was better than the eager approach and than standard solving strategy of \wasp. On the other hand, for high values of \textit{k} the advantages of the lazy approach disappear and the eager propagator obtained the best performance overall.
	Interestingly, \wasp without external propagators was better than its counterparts only when the value of \textit{k} was 95\%.
	Thus, the usage of external propagators was beneficial for solving this problem.
	
	\paragraph{Packing.}
	Concerning this problem, propagators were evaluated on 50 instances submitted to the Third ASP Competition.
	Interestingly, without the usage of external propagators none of the instances can be instantiated by the state of the art grounder \gringo, thus standard solving strategies were not feasible for this kind of problem.
	On the other hand, performance of propagators were much better. Indeed, lazy approach solved 20\% of the instances, while eager and post propagators were able to solve \textit{all} the considered instances.
	
	\paragraph{Natural Language Understanding (NLU).}
	Concerning NLU, propagators were evaluated on 50 instances using the objective functions proposed in~\cite{DBLP:journals/fuin/Schuller16}, namely \textit{Cardinality}, \textit{Coherence} and \textit{Weighted Abduction}.
	The performance of propagators was dependent on the specific objective function considered, e.g. the lazy approach was slightly faster than other propagators on \textit{Cardinality} and \textit{Weighted Abduction}, while for \textit{Coherence} post propagator was the best alternative.
	However, all the propagators outperform the standard solving strategy, thus showing also in this case a clear advantage of using custom propagators.
	
	\subsection{Heuristics}
	In \cite{DBLP:journals/tplp/DodaroGLMRS16}, the interface of \wasp was used for integrating domain-specific heuristics for two industrial problems proposed by Siemens, namely Partner Units and Combined Configuration.
	In the following, we review the main results obtained by \wasp in \cite{DBLP:journals/tplp/DodaroGLMRS16}.
	
	\paragraph{Partner Units.}	
	Concerning Partner Units problem, three different heuristics were proposed.
	All of them take advantage of $\mathit{AttachLiterals}$ to initialize their internal data structures and $\mathit{OnConflict}$ to update them during the search.
	In addition, method $\mathit{OnLiteralTrue}\ (\mathit{OnLiteralsTrue})$ and $\mathit{OnUnrollLiteral}$ are used to synchronize the internal state of the heuristics with the partial interpretation of the CDCL algorithm.
	Moreover, the heuristics are able to recognize that the current partial assignment cannot be completed to a valid solution.
	Thus, in this case, a restart is performed returning \textit{(restart)} after the subsequent call to the method $\mathit{SelectLiteral}$.
	The performance of \wasp employing such heuristics was empirically evaluated on 36 instances provided by Siemens.
	The results reported in \cite{DBLP:journals/tplp/DodaroGLMRS16} show that \wasp equipped with custom heuristics outperforms state-of-the-art approaches, including \clasp, \textsc{claspfolio} and \textsc{measp}.
	
	\paragraph{Combined Configurations.}
	Concerning Combined Configurations problems, a set of six heuristics were analyzed.
	All of them share a similar skeleton, and as for the Partner Units problem,
	use $\mathit{AttachLiterals}$ and $\mathit{OnConflict}$ to initialize and update the internal structures and strategies, respectively.
	Methods $\mathit{OnLiteralTrue}\ (\mathit{OnLiteralsTrue})$ and $\mathit{OnUnrollLiteral}$ are used to synchronize the internal state of the heuristics with the partial interpretation of the CDCL algorithm.
	An interesting aspect of some the strategies proposed is the interoperability with the default \minisat heuristic implemented in \wasp.
	Indeed, one strategy switches to the \minisat heuristic when some conditions are satisfied. This is implemented by returning $(\minisat,0)$ when the method $\mathit{SelectLiteral}$ is called.
	While, another strategy alternates the custom heuristic with the \minisat one. In particular, method $\mathit{SelectLiteral}$ returns $(\minisat, 1)$ every 10 seconds.
	The experiment considered 36 instances provided by Siemens.
	The results  in \cite{DBLP:journals/tplp/DodaroGLMRS16} show that \wasp equipped with the heuristic implementing the alternating strategy solves all the tested instances outperforming state-of-the-art approaches, including \clasp, \textsc{claspfolio} and \textsc{measp}.
	
	\section{Related Work} \label{sec:related}
	
	\paragraph{Propagators.}
	The extension of CDCL solvers with propagators is at the basis of Satisfiability Modulo Theories (SMT) solvers~\cite{DBLP:journals/jacm/NieuwenhuisOT06}.
	Indeed, external theories are usually implemented by means of propagators on top of state-of-the-art SAT solvers.
	Similar extensions have been envisaged also for ASP~\cite{DBLP:conf/ijcai/BartholomewL13}.
	Other extensions of ASP such as CASP~\cite{DBLP:conf/iclp/BaseliceBG05} or aggregates~\cite{DBLP:journals/tplp/AlvianoDM18} have been implemented by adding propagators to CDCL solvers \cite{DBLP:journals/tplp/OstrowskiS12}.
	
	The extension of \wasp presented in this paper can serve as a platform for implementing such language extensions. Indeed, new propagators can be added to implement specific constraints (such as \textit{acyclicity constraints}~\cite{DBLP:conf/lpnmr/BomansonGJKS15,DBLP:journals/fuin/BomansonGJKS16}), ASP modulo theories~\cite{DBLP:conf/ijcai/BartholomewL13,DBLP:conf/jelia/BartholomewL14} and CASP~\cite{DBLP:conf/iclp/BaseliceBG05}, and can be also used for boosting the performance of \wasp on specific benchmarks.
	
	An extension similar to the one presented in this paper has been implemented in solvers of Potassco project~\cite{DBLP:journals/aicom/GebserKKOSS11}.
	The ASP solver \clasp~\cite{DBLP:conf/lpnmr/GebserKK0S15} provides a C++ interface for post-propagation, where it is possible to invalidate a stable model candidate.
	The interface for defining new propagators is conceptually equivalent to the one presented in Section~\ref{sec:propagation}.
	However, at the moment \clasp does not support any external \textit{python} (or \textit{perl}) API to specify new propagators.
	A \textit{python} library is currently supported by \clingo~\cite{DBLP:journals/corr/GebserKKS14}.
	First versions of the API supported by \clingo (up to version 4)~\cite{DBLP:journals/corr/GebserKKS14} have no concept of post propagators but only support a function similar to
	\textit{CheckStableModel}, which is called whenever a stable model is found.
	The version 5 of \clingo~\cite{DBLP:conf/iclp/GebserKKOSW16} supports also a similar API to define external propagation using scripting languages.
	Several important differences exist between the interface of \wasp and the one of \clingo.
	In particular, \clingo provides only a post-propagator interface and no possibility for realizing an eager propagator (that runs before unit propagation is finished).
	Moreover, \wasp first collects constraints added in \textit{python} and then internally applies them and handles conflicts, while \clingo requires an explicit propagation call after each added constraint.
	Finally, \clingo offers the possibility to control both grounding and solving via \textit{python}, while \wasp only works on propositional programs.	
	
	\paragraph{Heuristics.}	
	A declarative approach to define domain specific heuristics in ASP is presented in~\cite{DBLP:conf/aaai/GebserKROSW13}. 
	The suggested \hclasp framework extends the \gringo language with a special \heuristic predicate. 
	Atoms over this predicate allow one to influence the choices made by the default heuristic of \clasp. 
	In fact, a user can provide initial weights, importance factors, sign selection and decision levels for atoms involved in non-deterministic decisions.
	The first three elements can be also specified in \wasp using methods $\mathit{Init}\minisat$, $\mathit{Factor}\minisat$ and $\mathit{Sign}\minisat$, respectively, while levels are not supported by the interface of \wasp.
	Moreover, \hclasp supports the definition of dynamic heuristics by considering only those atoms over \heuristic predicate that are true in the current interpretation. 
	In contrast to static heuristics, where heuristic decisions are encoded as facts, dynamic ones comprise normal rules with a \heuristic atom in the head.
	As argued in~\cite{DBLP:conf/lpnmr/GebserRS15}, the grounding of programs comprising definitions of dynamic heuristics can be expensive as it impacts on grounding speed and size. 
	The reason is that a grounder needs to output a rule for every possible heuristic decision.
	Since version 5 of \clingo, the \heuristic predicate has been replaced by the directive \texttt{\#heuristic}, whose working principles are similar.
	In \cite{DBLP:journals/aicom/Balduccini11}, a technique which allows learning of domain-specific heuristics in DPLL-based solvers is presented.
	The basic idea is to analyze off-line the behavior of the solver on representative instances from the domain to learn and use a heuristic in later runs. 
	The interface of \wasp presented in this paper could be considered for porting the ideas of \cite{DBLP:journals/aicom/Balduccini11} in a CDCL solver.	
	As far as we know, there are no other ASP solvers that support \textit{python} (or \textit{perl}) implementations to specify new heuristics.		
	
	\section{Conclusion}
	In this paper, we presented the external interface of the ASP solver \wasp conceived for easing the burden of extending its solving capabilities by means of new propagators and choice heuristics.
The implementation of the interface supports both rapid development by means of scripting languages, where no modifications to the solver is required, and performance-oriented development in \textit{C++}. 
Successful applications of the interface in the literature witness that the usage of the interface for developing domain-specific propagators and heuristics can be very effective for solving real-world problems and in general for speeding up the performance of \wasp. 

	\bibliographystyle{acmtrans}
	\bibliography{bibliography}	
	
	\label{lastpage}
\end{document}